\documentclass[twocolumn,times]{aastex63}

\usepackage{epstopdf,bm,amsmath}
\usepackage{orcidlink}
\usepackage[T1]{fontenc}
\usepackage{mathptmx}

\renewcommand{\vec}[1]{\bm{#1}}

\newcommand{\etal}{et al.}

\newcommand{\wind}{\emph{WIND}}

\newcommand{\aave}[1]{\left\langle #1\right\rangle} 

\usepackage[T1]{fontenc}
\usepackage{mathptmx}
\shorttitle{On the Statistics of Elsasser Increments in Solar Wind and MHD Turbulence}
\graphicspath{{./}{newplots/}}
\begin{document}

\title{On the Statistics of Elsasser Increments in Solar Wind and Magnetohydrodynamic Turbulence}

\author[0000-0002-3249-3335]{Juan C. Palacios}%\,\orcidlink{0000-0002-3249-3335}}
\affiliation{Florida Institute of Technology, 150 University Blvd, Melbourne, FL 32901, USA}
\author[0000-0002-2358-6628]{Sofiane Bourouaine}%\,\orcidlink{0000-0002-2358-6628}}
%\author{Sofiane Bourouaine\orcidlink{0000-0002-2358-6628}}%\,\orcidlink{0000-0002-2358-6628}}
\affiliation{Johns Hopkins University Applied Physics Laboratory, Laurel, MD, United States of America.}
\author[0000-0002-8841-6443]{Jean C. Perez}%\,\orcidlink{0000-0002-8841-6443}}
\affiliation{Florida Institute of Technology, 150 University Blvd, Melbourne, FL 32901, USA}

%\nocollaboration{2}

%% Note that the \and command from previous versions of AASTeX is now
%% depreciated in this version as it is no longer necessary. AASTeX 
%% automatically takes care of all commas and "and"s between authors names.

%% AASTeX 6.3 has the new \collaboration and \nocollaboration commands to
%% provide the collaboration status of a group of authors. These commands 
%% can be used either before or after the list of corresponding authors. The
%% argument for \collaboration is the collaboration identifier. Authors are
%% encouraged to surround collaboration identifiers with ()s. The 
%% \nocollaboration command takes no argument and exists to indicate that
%% the nearby authors are not part of surrounding collaborations.

%% Mark off the abstract in the ``abstract'' environment. 
%TC:break Abstract
\begin{abstract}
We investigate the dependency with scale of the empirical probability distribution functions (PDF) of Elsasser increments using large sets of WIND data (collected between 1995 and 2017) near 1 au. The empirical PDF are compared to the ones obtained from high-resolution numerical simulations of steadily driven, homogeneous Reduced MHD turbulence on a $2048^3$ rectangular mesh. A large statistical sample of Alfv\'enic increments is obtained by using conditional analysis based on the solar wind average properties. The PDF tails obtained from observations and numerical simulations are found to have exponential behavior in the inertial range, with an exponential decrement that satisfies power-laws of the form $\alpha_l\propto l^{-\mu}$, where $l$ the scale size, with $\mu$ around 0.2 for observations and 0.4 for simulations. PDF tails were extrapolated assuming their exponential behavior extends to arbitrarily large increments in order to determine structure function scaling laws at very high orders. Our results points to potentially universal scaling laws governing the PDF of Elsasser increments and to an alternative methodology to investigate high-order statistics in solar wind observations.
\end{abstract}

%TC:break _main_
%% Keywords should appear after the \end{abstract} command. 
%% See the online documentation for the full list of available subject
%% keywords and the rules for their use.

%% From the front matter, we move on to the body of the paper.
%% Sections are demarcated by \section and \subsection, respectively.
%% Observe the use of the LaTeX \label
%% command after the \subsection to give a symbolic KEY to the
%% subsection for cross-referencing in a \ref command.
%% You can use LaTeX's \ref and \label commands to keep track of
%% cross-references to sections, equations, tables, and figures.
%% That way, if you change the order of any elements, LaTeX will
%% automatically renumber them.
%%
%% We recommend that authors also use the natbib \citep
%% and \citet commands to identify citations.  The citations are
%% tied to the reference list via symbolic KEYs. The KEY corresponds
%% to the KEY in the \bibitem in the reference list below. 

\section{Introduction}\label{sec:intro}
Decades of spacecraft observations have shown that the solar wind properties exhibit random fluctuations over a wide range of lengthscales consistent with a turbulent state~\citep{bruno13}. 
For scales larger than any plasma microscale, such as the ion and electron gyroradius, the power spectra of velocity and magnetic fluctuations obeys a power law similar to the Kolmogorov $k^{-5/3}$ law (K41 hereafter) for fluid turbulence~\citep{kolmogorov41,kolmogorov41a}, which has long been thought to arise from an incompressible Magnetohydrodynamics (MHD) turbulent cascade mediated by Alfv\'enic fluctuations~\citep{coleman68,belcher69}. 

Since the pioneering work of \cite{iroshnikov63,iroshnikov64} and \cite{kraichnan65} (IK hereafter), predicting a power spectrum scaling $\propto k^{-3/2}$, most MHD turbulence models~\citep{goldreich95,lithwick07,chandran08,beresnyak08,boldyrev05,perez09} are based on Kolmogorov's assumption that the spatial distribution of fluctuations is self similar in the \emph{inertial range}. 
Self-similarity is intuitively associated with the fact that fluctuations at each scale $l\sim 1/k$ are space-filling. However, fluid turbulence experiments and simulations~\citep{anselmet84,gotoh02}, and solar wind observations~\citep{burlaga91} show that at smaller scales the distribution of turbulent fluctuations becomes increasingly sparse. This departure from self-similarity, which is called intermittency, plays an important role in plasma heating processes~\citep{sundkvist07,zhdankin16}.

The first observations of intermittency in the solar wind by~\cite{burlaga91} were followed by numerous works on the subject using nearly every spacecraft to date, for a recent review see~\cite{bruno19}. The large majority of these works have focused on the intermittency of velocity $\vec v$ and magnetic field $\vec B$. However, the so-called Elsasser fields $\vec z^\pm=\vec v\pm\vec B/\sqrt{4\pi\rho}$ are more fundamental variables to study MHD turbulence given that, contrary to kinetic and magnetic energy, their energies are subject to a conservative cascade. 
In this work we use the largest statistical sample to date of turbulent increments in the solar wind from the WIND spacecraft, spanning 23 years from 1995 to 2017, and from high-resolution numerical simulations of steadily-driven MHD turbulence to investigate the scale-dependent Probability Distribution Functions (PDF) of Elsasser increments. Our analysis is based on conditional statistics to ensure Elsasser increments belong to Alfv\'enic fluctuations. The large statistical sample allows us to empirically estimate the PDF of the turbulence increments over many standard deviations, capturing significant portion of those heavy tails that are signature of intermittency. Exponential least-square-fits of these tails are obtained to investigate their dependency on the scale, which in turn we use to extrapolate empirical PDF to obtain estimates of structure functions to higher orders than those allowed by the finite data sample. These results show the first empirical evidence of potentially universal scaling laws governing the PDF tails of Elsasser increments in the solar wind, and enable a new venue to investigate intermittency that allows for direct comparisons with new and existing theoretical models.  This paper is organized as follows. In section~\ref{sec:theory} we provide a brief theoretical background of intermittency in MHD and solar wind turbulence in order to provide minimal context and notation that will be used in the rest of this paper. In section~\ref{sec:data} we describe the solar wind observations from the~\wind~spacecraft and numerical simulations, as well as the methodology used in this work to obtain Elsasser increments and their corresponding PDF. In section~\ref{sec:pdf} we show and discuss our results and in section~\ref{sec:conclusions} we present our conclusions.

\section{Theoretical framework}\label{sec:theory}
The statistical properties of MHD turbulence are often described in terms of the statistical moments of longitudinal increments~\citep{Biskamp00}
\begin{equation}
    \delta z^\pm_L(\vec l,\vec x,t)=\hat{\vec l}\cdot\delta\vec z^\pm(\vec l,\vec x,t),\label{eqn:inc}
\end{equation}
where $\delta\vec z^\pm(\vec l,\vec x,t)\equiv\vec z^\pm(\vec x+\vec l/2,t)-\vec z^\pm(\vec x-\vec l/2,t)$ represent a typical turbulent fluctuation at scale $l=|\vec l|$, $\vec l$ represents a scale vector in the plane perpendicular to the mean background magnetic field,  and $\hat{\vec l}$ is the unit vector in the direction of $\vec l$. In the homogeneous and stationary state, these moments, known as \emph{structure functions}, are expected to satisfy universal power laws of the form
\begin{equation} 
S_n^\pm(l)=\langle |\delta z^\pm_L(\vec x,\vec l)|^n\rangle=a_n^\pm l^{\zeta_n^\pm},\label{eqn:sfe}
\end{equation}
for length-scales $l$ in the inertial range. In this expression, we use $\aave{\cdots}$ to denote a suitable ensemble average and $a_n^\pm$ are non-universal coefficients that solely depend on $n$. 

Kolmogorov's self-similarity assumption implies that the scaling exponents $\zeta_n^\pm$ are linear in $n$
\begin{equation}
    \zeta_n^\pm=h^\pm n, \quad h>0,
    \label{eqn:zetaSS}
\end{equation}
where $h^\pm$ are constants. Although dimensional arguments can be used to uniquely determine $h=1/3$ in fluids, they alone are not sufficient to determine $h^\pm$ in MHD. IK and K41 scaling correspond to $h^\pm=1/4$ and $h^\pm=1/3$, respectively.

Multifractality arises assuming that the turbulence cascade accumulates on multiple fractal sets with different fractal dimensions, resulting in a range of scaling exponents $h\in[h_{\rm min},h_{\rm max}]$ in which fluctuations satisfy local scale invariance with the corresponding scaling index $h$. In this picture, the scaling of structure functions at each order $n$ results from the fractal set with index $h$ that has the most dominant contribution to the average of the $n^{\rm th}$ order power of the corresponding increment. As a result, the dependency with $n$ of the scaling exponents $\zeta_n$ becomes non-linear, because at each $n$ the largest contribution to the statistical average arises from a fractal set with a different $h$ value.

Structure functions are defined in terms of a hypothetical ensemble averages, which assume an arbitrarily large number of identical realizations of the system. In practice, an ergodicity assumption has to be invoked in the homogeneous and stationary state in order to empirically estimate these averages using a finite sample. In reality, an exact calculation of these structure functions is only possible if the PDF governing the increments were known, which is arguably the holy grail of turbulence theory. The PDF $\mathcal P(u)$ of a random variable $u$ is defined such that $\mathcal P(u)du$ is the probability of finding the random variable $u$ between $u$ and $u+du$. The structure functions can then be written in terms of the PDF $\mathcal P^\pm(u,l)$ associated with the corresponding Elsasser fields  as
\begin{equation}
    S_n^\pm(l) = \int_{-\infty}^\infty u^n\mathcal P^\pm(u,l)du.
    \label{eqn:sft}
\end{equation}
where  $u$ represents the increments of the fields. Note that in this work we define structure functions in terms of the increment magnitude $|u|$, which in general exhibit a more distinct scaling behavior and are expected to show the same exponents~\citep{Biskamp00}. Performing a statistical study of increments using the PDF is an alternative venue to study intermittency in the inertial range \citep{sorriso-valvo99,barndorff-nielsen04}, which is one of the main objectives of this work.

\begin{figure*}[!ht]
\plotone{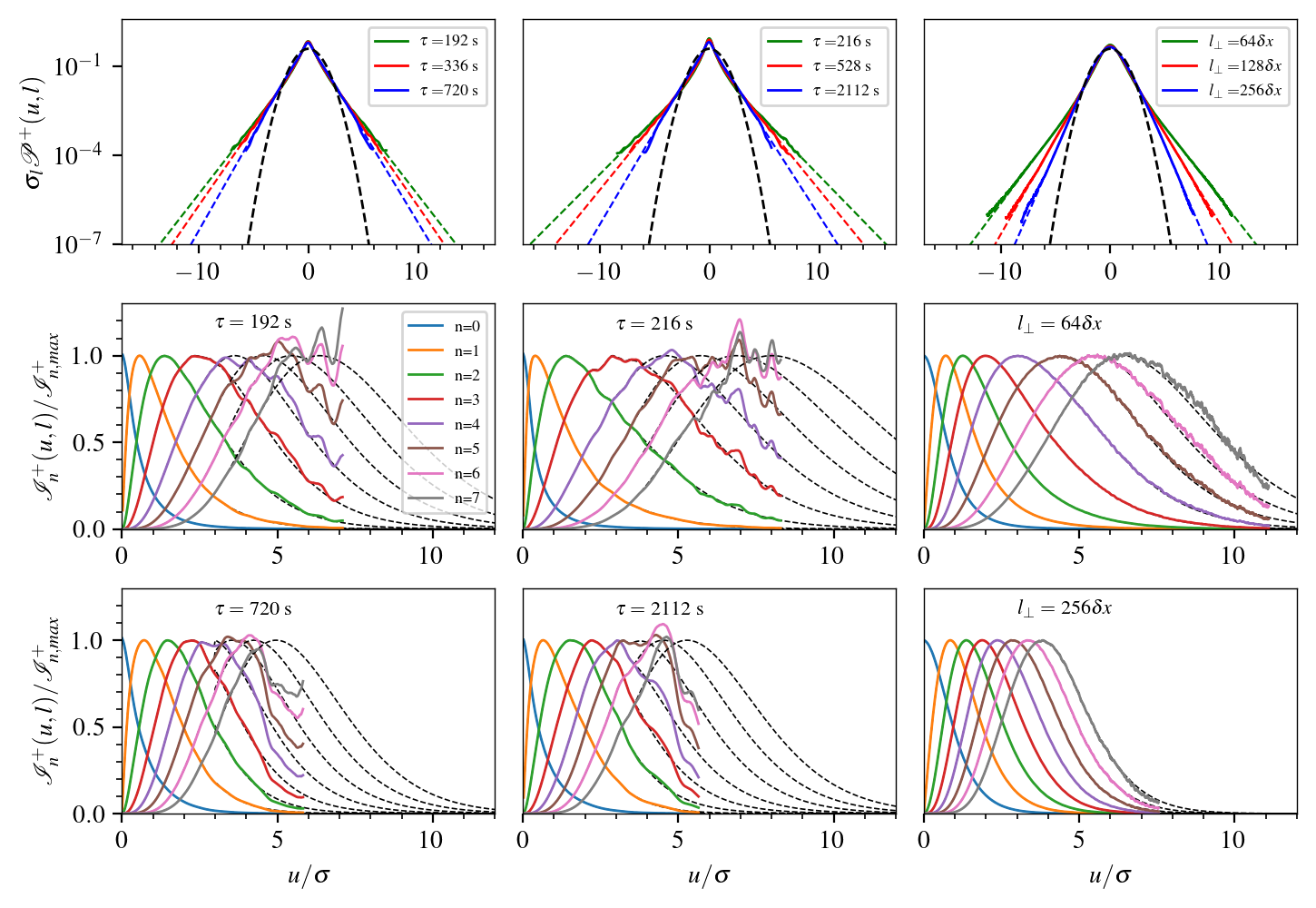}
\caption{Top panels: continuous lines represent the estimated Probability Distribution Functions $\mathcal{P}^+(u, l)$ versus $u/\sigma_l$ for fast wind (left), slow wind (middle) and simulation (right), where $u=\delta z_L^+$. Note that for spacecraft data the space lag is $l=V_{SW}\tau$ via Taylor's Hypothesis, and $\sigma_l$ represents the standard deviation at scale $l$. Middle and bottom panels:  distribution of $n^{\rm th}$ moment $\mathcal I^+_n(u,l)$ normalized to its maximum value ($\mathcal I_{n,max}^+$) for fast wind (left), slow wind (middle) and simulation (right) for two different lags near the end (middle panel) and beginning (bottom panel) of the inertial range. Dashed lines represents the approximation of the PDF extending the tails assuming an exponential approximation.}
\label{fig:pdfs}
\end{figure*}

\section{Data and methodology} \label{sec:data}

We use data for density $\rho$, magnetic field $\vec B$ and velocity $\vec v$ from the {\it WIND} spacecraft. Twenty three (23) years of data from WIND/3DP instrument (3D plasma analyzer) with resolution of $\sim 24$~s were used.  Velocity and magnetic field increments were carefully selected to ensure they belong to periods of homogeneous and incompressible turbulence in the slow and fast solar wind. For solar wind observations data were first resampled to a uniform grid of 24~s using linear interpolation and any gaps that the data may have were discarded.  The local mean quantities $\rho_{avg}(t)$, $\vec B_{avg}(t)$ and $\vec v_{avg}(t)$ were calculated for each point using a moving average with a two-hour window for fast wind and an eight-hour window for slow wind. In order to match the best possible conditions for Alfv\'enic turbulence, the mean plasma properties were restricted so that the mean bulk speed remains in the range $500 < v_{avg} < 700$~km/s for fast wind and $280 < v_{avg} < 480$~km/s for slow wind, $\delta B/B_{avg}\le0.2$ with mean magnetic field $B_{avg} < 12$~nT and $B_{avg} < 8$~nT for fast and slow wind, respectively, and density fluctuations are much smaller than the local average density, $\delta \rho/\rho \le 0.15$ to ensure incompressibility. Assuming the turbulence is strong~\citep{goldreich95}, we estimate the turbulence anisotropy $k_\|/k_\perp\sim\delta B/B_{ave}\sim0.2$, and restrict the sampling to be nearly perpendicular to the background field, i.e. $\sin\theta_{VB}\gg0.2$, where $\theta_{VB}$ is the angle between $\vec v_{avg}$ and $\vec B_{avg}$. Based on this condition we restrict the sampling angle to be in the range $50^\circ\le\theta_{VB}\le 130^\circ$.

We invoke Taylor frozen-in-flow hypothesis \citep{Taylor38} to interpret temporal signals as spatial variations, where the correspondence between spatial increments at scale $l$ correspond to temporal increments at scale $\tau$, where $\tau=l/v_{sw}$ and $v_{sw}$ is the mean solar wind speed. Based on this assumption, Equation \eqref{eqn:inc} can be used to calculate increments whenever the conditions described above (to ensure the increment belongs to an Alfv\'enic interval) are satisfied at the three times, $t-\tau/2$, $t$ and $t+\tau/2$. Using data from 1995 to 2017, around $1.5 \times 10^6$ realizations for fast wind and $1 \times 10^6$ realizations for slow wind were obtained.

In order to establish comparisons with observations, we also use pseudo-spectral simulations of steadily-driven, strong balanced Reduced MHD turbulence (RMHD) on a rectangular grid with of $2048^3$ mesh points, which are described in detail in~\cite{perez12}. The simulations describe turbulent Alfv\'enic fluctuations like those we focus on in observations, with the exception that simulations have zero cross helicity (balanced turbulence or $z^+\sim z^-$).  A total of $30$ snapshots of the turbulent fields $\vec z_\alpha(\vec x)=\vec z^\pm(\vec x, t_\alpha)$ with $\alpha=1,2,...,30$ in the steady state are used from the simulations.  Field increments perpendicular to the background of the magnetic field are sampled at a random set of $N$ points $\vec x_i$, with $i=1,2,...,N$, generating around $2 \times 10^9$ realizations.

Once increments are calculated for these three systems, empirical PDF of Elsasser increments are constructed from estimated histograms of the statistical samples for each time scale $\tau$ in observations.

\begin{figure*}[!ht]
\plotone{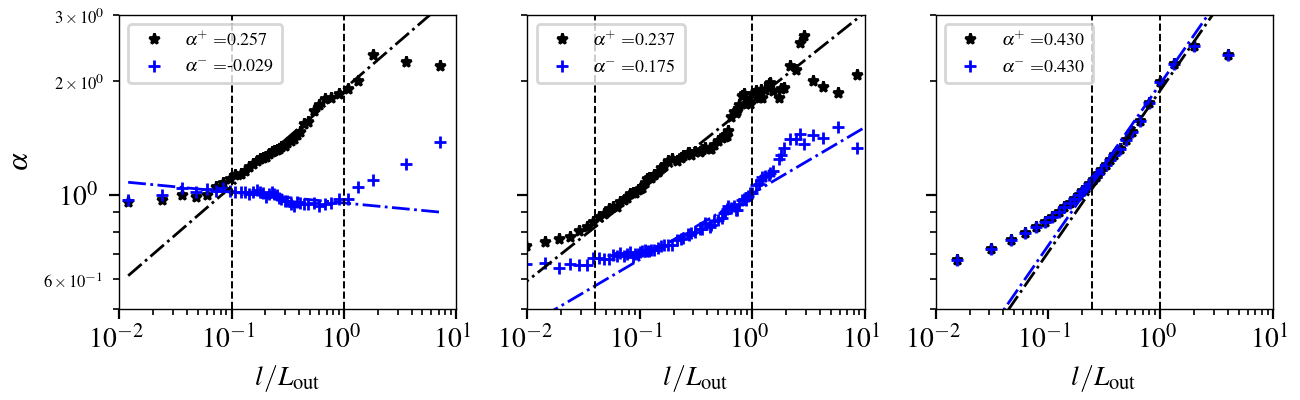}
\caption{Exponential decrements of extended tails $\alpha_l$ for $\delta z^+_l$ (black) and $\delta z^-_l$ (blue) in fast solar wind (left), slow solar wind (center) and simulation (right). Here $l/L_{\rm out}$ represents the increment scale normalized with respect to the outer scale $L_{\rm out}$ which correspond to 700~s for fast wind, 2000~s for slow wind and $256\delta x$ for simulations. 
%\color{blue}{The $x$ axis is normalized to the outer scale, so $x=l/l_{out}$.  
Vertical lines represents the limits of the inertial range.}
\label{fig:alpha}
\end{figure*}

\section{Results and Discussions}\label{sec:pdf}

Solid lines in Figure \ref{fig:pdfs} show the estimated PDF for the three systems, normalized using the scale-dependent standard deviation $\sigma_l$, for three representative scales within the inertial-range: near the outer scale (where energy is injected) in green, near the middle of the inertial range in red, and near the dissipation range in blue, while black curves represent a normalized Gaussian for reference.  Dashed lines represent an exponential extension of these PDF. The presence of intermittency becomes evident as the PDF tails become heavier at smaller scales, signifying a departure from self-similarity.

One of the main challenges in the empirical estimation of PDF is that any measurement necessarily involves a finite number of samples, leading to noisy tails, which in turn becomes a source of error in the estimation of statistical averages. In order to reduce the noise, we applied a \cite{savitzky64} filter based on third-order polynomials to each PDF. From our data samples, the resulting PDF cover an increment range of up to seven standard deviations for fast wind, ten for slow wind and fifteen for simulations, making them to the best of our knowledge the largest well defined tails to date~\citep{sorriso-valvo99, bruno99, sundkvist07, salem09, greco10, alexandrova13}.

Most empirical results obtained in simulations~\citep{Chandran15a, mallet16} and observations~\citep{macbride05,bruno19} thus far are only valid for structure functions of low order, i.e., for small values of $n\lesssim 4$ or $5$.  At a given order $n$ and scale $l$, $S^\pm_n(l)$ represents the area under the curve generated by $\mathcal I_n^\pm(u,l)=u^n\mathcal P^\pm(u,l)$, which for convenience we call the distribution of $n^{\rm th}$ order moments. Empirical estimations of $\mathcal I_n^\pm(u,l)$ for solar wind observations show that the distribution does not drop to zero fast enough, within the range of measured values, at high order (colored curves in middle and lower panels of Figure \ref{fig:pdfs}), resulting in an underestimation of the corresponding moment, and providing just the first three or four moments with reasonable accuracy. If one assumes that the observed exponential behavior of the tails in the inertial range persists for large increment values (or rare events not captured by the original data), we can use least-fit square (from $3\sigma$, $3.5\sigma$ and $4.5\sigma$ for fast wind, slow wind and simulations, respectively) to analytically extrapolate the tails as $\propto e^{-\alpha_l|l|}$, where $\alpha_l$ is a scale-dependent free parameter describing the tails' exponential decrement\footnote{In this work we assume the skewness of the PDF is small enough to assume that the tails are nearly symmetric}. 
%\textcolor{blue}{Tails were extrapolated until the integrand in equation \eqref{eqn:sft} drops to zero.}

Interestingly, the exponential decrements $\alpha_l^\pm$, shown in Figure~\ref{fig:alpha}, exhibit a power law behavior in the inertial range (indicated by the vertical dashed lines) both in observations and simulations. For fast and slow wind, the scaling exponent for $\alpha_l^+$ are remarkably similar $\alpha_l^+\propto l^{-0.25}$ for $\vec z^+$ increments, while the scaling for $\vec z^-$ increments differs between fast and slow wind. In the former case, it is found that $\alpha_l^-$ is nearly scale-independent, suggesting that $\vec z^-$ increments are less intermittent. For slow wind, $\alpha_l^-$ power law is slightly flatter than that of $\alpha_l^+$, suggesting that both $\vec z^+$ and $\vec z^-$ exhibit intermittent behavior. The power laws that we obtained for the exponential decrement $\alpha_l^\pm$ in solar wind measurements are remarkably close to those previously reported in hydrodynamic turbulence from wind tunnel experiments, $\alpha_l\propto l^{-0.17}$, by~\cite{praskovsky94}.

\begin{figure}[!t]
\plotone{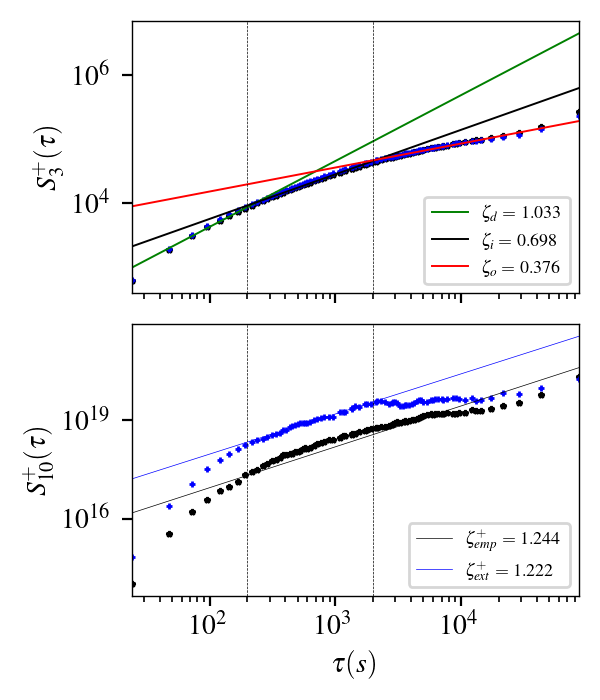}
\caption{Third (top) and tenth (bottom) order structure functions for $\delta z^+$ of slow wind calculated using the empirical PDF (black marks) and the extrapolated PDF using exponential tails (blue marks) versus time lags. Green and red lines show power-law fits for the dissipative and injection regions, respectively, while the black line indicate the corresponds to the inertial range.}
\label{fig:sf3}
\end{figure}

In order to obtain better estimates of $S_{n}^\pm(l)$ at high order, under the assumption that the exponential tails extend beyond the measurable range, we use PDF with extrapolated tails, shown by the dashed lines in Figure \ref{fig:pdfs}, to numerically evaluate the integral in equation \eqref{eqn:sft}. Figure \ref{fig:sf3} shows $S_n^+(\tau)$ for $n=3$ and $n=10$ for slow wind calculated using the PDF directly constructed from observations (black stars) and PDF with extrapolated exponential tails (blue stars). The overlap between black and blue stars in the top panel shows, as expected, that using either PDF to estimate $S_3$ leads to the same result. The bottom panel shows that $S_{10}$ is underestimated at each scale when the empirical PDF is used vs the extrapolated one. Remarkably, its inertial-range scaling remains nearly identical for both PDF, suggesting that at each scale $S_{10}$ estimated from the empirical PDF represents the same fraction of its estimate with the extrapolated one. A similar result was observed in neutral fluid experiments by~\cite{anselmet84} up to order $n=18$. 
\begin{figure*}[!ht]
\plotone{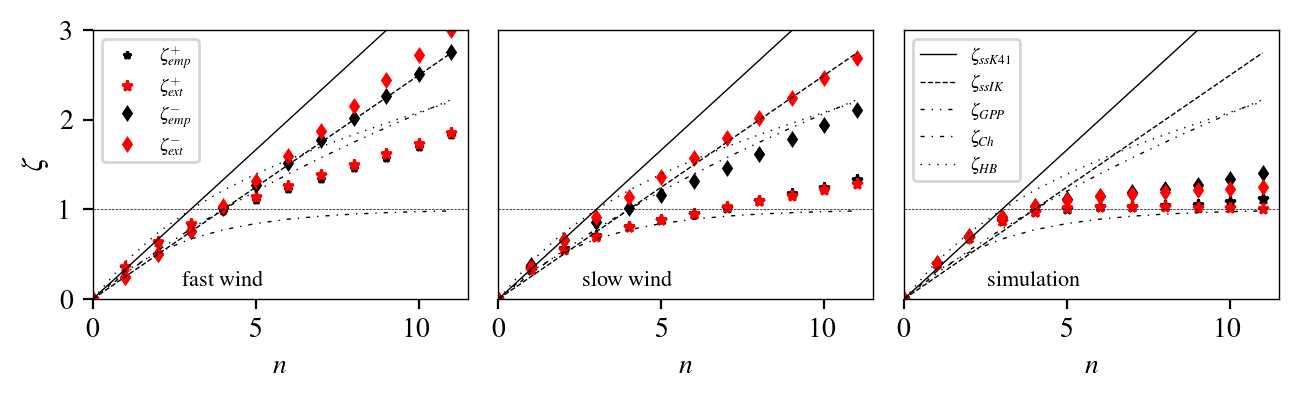}
\caption{Scaling exponents for $\delta z_L^+$ and $\delta z_L^-$ increments for fast wind (left), slow wind (middle) and simulation (right) directly calculated with the experimental data ($\zeta_e$) and using the PDF modeled with exponential tails ($\zeta_m$). Extended Self Similarity is applied in all the cases.}
\label{fig:scal}
\end{figure*}

We identify the inertial range as the region where its power-law fit (black line) intersect the corresponding fits in the dissipation (green line) and outer scale (red line) ranges. Using this method we identify the inertial range from 200~s to 2000~s for slow wind, from 200~s to 700~s for fast wind, and $64\delta x$ to $256\delta x$ where $\delta x$ represents the grid size in simulations.  In order to determine the scaling exponents more accurately we use the so-called Extended Self Similarity (ESS)~\citep{Benzi93}.  For reasons that are still not well understood, plotting $S_n^\pm$ as a function of $S_3^\pm$ instead of $l$ leads to extended power-law scaling, even outside the inertial range. The reason why~\cite{Benzi93} selected $S_3$ is because this is proportional to the scale-size in hydrodynamics.  Although for MHD turbulence $S_3^\pm$ is not proportional to $l$, it has also been found that using ESS  leads to extended power law regions. The only disadvantage is that scaling exponents $\gamma_n^\pm$ measured from ESS are related to $\zeta_n^\pm$ as $\gamma_n^\pm=\zeta_n^\pm/\zeta_3^\pm$, which still requires an accurate estimation of $\zeta_3^\pm$.

Figure \ref{fig:scal} shows the scaling exponents of structure functions up to order $n=12$ for $\delta z^+$ and $\delta z^-$ using ESS for the three systems, namely, the fast and slow wind as well as simulations. The scaling exponents $\zeta_n^+$ are represented by black symbols and $\zeta_n^-$ by red symbols, while those exponents resulting from the empirical PDF are represented by star-symbols while those resulting from the empirical PDF are represented by diamonds. With the exception of $\zeta_n^-$ in fast wind (left panel) all scaling exponents are non-linear in $n$, which suggest multifractal behavior.  

One of the most successful multifractal models in hydrodynamic turbulence was presented by~\cite{she94}, based in Log-Poisson distributions, which was later extended by \cite{horbury97} to MHD turbulence for K41 scaling ($h=1/3$)
\begin{equation}
    \zeta_n^{\rm HB}=\frac{n}{9}+1-\left(\frac{1}{3}\right)^{n/3}.
    \label{eqn:zetaHB}
\end{equation}
and by \cite{grauer94} and \cite{politano95} for the IK scaling ($h=1/4$)
\begin{equation}
    \zeta_n^{\rm GPP}=\frac{n}{8}+1-\left(\frac{1}{2}\right)^{n/4}.
    \label{eqn:zetaGPP}
\end{equation}
These models are of particular interest because they contain no freely adjustable parameters. 
More recently, \cite{Chandran15a} proposed a new model based on Alfv\'en-wave collisions, assuming that each balanced collision reduces a fluctuation's amplitude by a constant factor $\beta\simeq0.691$, leading to a simple relation for the scaling exponents, 
\begin{equation}
    \zeta_n^{\rm CH}=1-\beta^n.
    \label{eqn:zetaCh}
\end{equation}
A similar result was obtained by~\cite{Mallet17a} phenomenological model with $\beta=0.7$. 

Figure~\ref{fig:scal} shows comparisons between empirical results and the following theoretical predictions: the self-similar models with $h^\pm=1/3$ (ssK41) and $h^\pm=1/4$ (ssIK) in equation \eqref{eqn:zetaSS}, Horbury and Balogh (HB), Grauer \etal, Politano and Pouquet (GPP) and Chandran \etal~(CH) models in equations \eqref{eqn:zetaHB}, \eqref{eqn:zetaGPP} and \eqref{eqn:zetaCh} respectively.  For $\zeta^+_n$, we observe that up to $n=12$ the values estimated from the empirical PDF (or the standard method) are very close to those obtained with extrapolated PDF. The multifractal behavior of $\zeta_n^+$ is evident in the strong departure from self-similarity in all three cases.  In contrast, $\zeta_n^-$ shows little departure from self-similarity and very close to the ssIK model calculated using PDFs with extrapolated tail. As opposed to $\zeta_n^+$, our results also show that for high orders the scaling exponents $\zeta_n^-$ are in fact different when the PDF tails are extrapolated. The exponent of $S_4^+$ for fast wind is very close to one, suggesting that $\langle \delta z^+\rangle \propto l$, consistent IK scaling. For slow wind, the first six scaling exponents are remarkably close to CH model. In contrast to observations, $\zeta^+$ and $\zeta^-$ are very similar in simulations because the they correspond to a steady state of balanced turbulence, in which Elsasser variables have comparable amplitudes.

\section{Conclusions}\label{sec:conclusions}

Accurate measurements of structure functions provide critical information about the development of intermittency and help to understand the energy transfer in the inertial range.  In this work we proposed a methodology to calculate Elsasser increments that allows us to collect the largest possible statistics of Alfv\'enic solar wind for fast and slow wind. The statistics was large enough that we were able to construct PDF of increments at each scale spanning up to seven and ten standard deviations for fast and slow wind, respectively, with less noisy and statistically better defined tails that previous works~\citep{sorriso-valvo99, bruno99, sundkvist07, salem09, greco10,  alexandrova13, osman14}.

Proper estimation of high order structure functions requires accurate estimations of the PDF tails.  However, this is normally not possible with a finite statistical sample, as they rely on rare events. Because of our substantially large statistical sample obtained through conditioning of 23 years of observations ($\sim10^6$ samples) and even larger sample in simulations ($\sim 10^9$), we were able to identify exponential tail behavior over several standard deviations in the inertial range, in which the exponential decrement satisfies well-defined power-law behavior of the form $\alpha_l\propto l^{-\mu}$, with $\mu$ around 0.2 for observations and 0.4 for simulations. This observed scaling of the exponential decrement, not previously reported in the solar wind literature, is very similar to those observed in fluid experiments, suggesting that this is potentially an intrinsic (or universal) property of the PDF of Elsasser increments. If this exponential behavior persists well beyond the observed range, it could help us obtain a deeper understanding of intermittency in the solar wind, such as scaling of high-order structure functions, beyond the limit imposed by empirical data with finite sample.

Under the assumption that in the inertial range the behavior of the tails remains exponential beyond the maximum measurable increment, as observed in simulations, we extrapolated the PDFs as long as needed to improve calculations of structure functions and the corresponding scaling exponents.  The scaling exponent of $S_3$ in the inertial range is observed to be smaller than unity for both Elsasser increments in the three experiments, suggesting a deviation from K41 theory and similar models. The scaling exponents $\zeta^+$ confirm the multifractal nature of $\delta z^+$ increments.  Although none of the models presented in section \ref{sec:pdf} fully describes the behavior of exponents for $\delta z^+$ in all three systems, we found that for fast wind observations $S_4$ has a value very close to one, consistent with IK scaling, while for slow wind, the first moments are very close to those predicted by CH model, and substantially deviates from observations for $n > 6$. For $\delta z^-$, both fast and slow wind reveal self-similar behavior with $h=1/4$ using PDFs with extrapolated tails, which can potentially shed some light about the possible sources of these fluctuations.

%\section{Acknowledgements}
\acknowledgements

This work was supported from NASA-NNX16AH92G and NSF-SHINE-AGS-1752827 grants. High-performance computing (HPC) resources were provided by the Argonne Leadership Computing Facility (ALCF) at Argonne National Laboratory, which is supported by the  U.S. Department of Energy under contract No. DE-AC02-06CH11357. HPC resources were also provided by the Texas Advanced Computing Center (TACC) at the University of Texas at Austin, NSF-XSEDE Project No. TG-ATM100031, and Blueshark at the Florida Institute of Technology supported by NSF-CNS-09-23050 grant.

%\bibliography{Intermittency2.bib}

\begin{thebibliography}{}
\expandafter\ifx\csname natexlab\endcsname\relax\def\natexlab#1{#1}\fi
\providecommand{\url}[1]{\href{#1}{#1}}
\providecommand{\dodoi}[1]{doi:~\href{http://doi.org/#1}{\nolinkurl{#1}}}
\providecommand{\doeprint}[1]{\href{http://ascl.net/#1}{\nolinkurl{http://ascl.net/#1}}}
\providecommand{\doarXiv}[1]{\href{https://arxiv.org/abs/#1}{\nolinkurl{https://arxiv.org/abs/#1}}}

\bibitem[{Alexandrova {et~al.}(2013)Alexandrova, Chen, Sorriso-Valvo, Horbury,
  \& Bale}]{alexandrova13}
Alexandrova, O., Chen, C. H.~K., Sorriso-Valvo, L., Horbury, T.~S., \& Bale,
  S.~D. 2013, Space Science Reviews, 178, 101,
  \dodoi{10.1007/s11214-013-0004-8}

\bibitem[{Anselmet {et~al.}(1984)Anselmet, Gagne, Hopfinger, \&
  Antonia}]{anselmet84}
Anselmet, F., Gagne, Y., Hopfinger, E.~J., \& Antonia, R.~A. 1984, Journal of
  Fluid Mechanics, 140, 63, \dodoi{10.1017/S0022112084000513}

\bibitem[{{Barndorff-Nielsen} {et~al.}(2004){Barndorff-Nielsen}, Bl{\ae}sild,
  \& Schmiegel}]{barndorff-nielsen04}
{Barndorff-Nielsen}, O.~E., Bl{\ae}sild, P., \& Schmiegel, J. 2004, The
  European Physical Journal B - Condensed Matter and Complex Systems, 41, 345,
  \dodoi{10.1140/epjb/e2004-00328-1}

\bibitem[{Belcher {et~al.}(1969)Belcher, Davis, \& Smith}]{belcher69}
Belcher, J.~W., Davis, Jr., L., \& Smith, E.~J. 1969, Journal of Geophysical
  Research, 74, 2302, \dodoi{10.1029/JA074i009p02302}

\bibitem[{Benzi {et~al.}(1993)Benzi, Ciliberto, Tripiccione, Baudet, Massaioli,
  \& Succi}]{Benzi93}
Benzi, R., Ciliberto, S., Tripiccione, R., {et~al.} 1993, Physical Review E,
  48, R29, \dodoi{10.1103/PhysRevE.48.R29}

\bibitem[{Beresnyak \& Lazarian(2008)}]{beresnyak08}
Beresnyak, A., \& Lazarian, A. 2008, The Astrophysical Journal, 682, 1070,
  \dodoi{10.1086/589428}

\bibitem[{Biskamp \& M{\"u}ller(2000)}]{Biskamp00}
Biskamp, D., \& M{\"u}ller, W.-C. 2000, Physics of Plasmas, 7, 4889,
  \dodoi{10.1063/1.1322562}

\bibitem[{Boldyrev(2005)}]{boldyrev05}
Boldyrev, S. 2005, The Astrophysical Journal Letters, 626, L37,
  \dodoi{10.1086/431649}

\bibitem[{Bruno(2019)}]{bruno19}
Bruno, R. 2019, Earth and Space Science, 6, 656, \dodoi{10.1029/2018EA000535}

\bibitem[{Bruno {et~al.}(1999)Bruno, Bavassano, Pietropaolo, Carbone, \&
  Veltri}]{bruno99}
Bruno, R., Bavassano, B., Pietropaolo, E., Carbone, V., \& Veltri, P. 1999,
  Geophysical Research Letters, 26, 3185, \dodoi{10.1029/1999GL010668}

\bibitem[{Bruno \& Carbone(2013)}]{bruno13}
Bruno, R., \& Carbone, V. 2013, Living Reviews in Solar Physics, 10, 2,
  \dodoi{10.12942/lrsp-2013-2}

\bibitem[{Burlaga(1991)}]{burlaga91}
Burlaga, L.~F. 1991, Journal of Geophysical Research: Space Physics, 96, 5847,
  \dodoi{10.1029/91JA00087}

\bibitem[{Chandran(2008)}]{chandran08}
Chandran, B. D.~G. 2008, The Astrophysical Journal, 685, 646,
  \dodoi{10.1086/589432}

\bibitem[{Chandran {et~al.}(2015)Chandran, Schekochihin, \&
  Mallet}]{Chandran15a}
Chandran, B. D.~G., Schekochihin, A.~A., \& Mallet, A. 2015, The Astrophysical
  Journal, 807, 39, \dodoi{10.1088/0004-637X/807/1/39}

\bibitem[{Coleman(1968)}]{coleman68}
Coleman, Jr., P.~J. 1968, The Astrophysical Journal, 153, 371,
  \dodoi{10.1086/149674}

\bibitem[{Goldreich \& Sridhar(1995)}]{goldreich95}
Goldreich, P., \& Sridhar, S. 1995, The Astrophysical Journal, 438, 763,
  \dodoi{10.1086/175121}

\bibitem[{Gotoh {et~al.}(2002)Gotoh, Fukayama, \& Nakano}]{gotoh02}
Gotoh, T., Fukayama, D., \& Nakano, T. 2002, Physics of Fluids, 14, 1065,
  \dodoi{10.1063/1.1448296}

\bibitem[{Grauer {et~al.}(1994)Grauer, Krug, \& Marliani}]{grauer94}
Grauer, R., Krug, J., \& Marliani, C. 1994, Physics Letters A, 195, 335,
  \dodoi{10.1016/0375-9601(94)90038-8}

\bibitem[{Greco {et~al.}(2010)Greco, Servidio, Matthaeus, \& Dmitruk}]{greco10}
Greco, A., Servidio, S., Matthaeus, W., \& Dmitruk, P. 2010, Planetary and
  Space Science, 58, 1895, \dodoi{https://doi.org/10.1016/j.pss.2010.08.019}

\bibitem[{Horbury \& Balogh(1997)}]{horbury97}
Horbury, T.~S., \& Balogh, A. 1997, Nonlinear Processes in Geophysics, 4, 185,
  \dodoi{10.5194/npg-4-185-1997}

\bibitem[{Iroshnikov(1963)}]{iroshnikov63}
Iroshnikov, P.~S. 1963, Astronomicheskii Zhurnal, 40, 742.
\newblock \url{http://adsabs.harvard.edu/abs/1963AZh....40..742I}

\bibitem[{Iroshnikov(1964)}]{iroshnikov64}
---. 1964, Soviet Astronomy, 7, 566.
\newblock \url{http://adsabs.harvard.edu/abs/1964SvA.....7..566I}

\bibitem[{Kolmogorov(1941{\natexlab{a}})}]{kolmogorov41}
Kolmogorov, A.~N. 1941{\natexlab{a}}, Proceedings: Mathematical and Physical
  Sciences, 434, 15

\bibitem[{Kolmogorov(1941{\natexlab{b}})}]{kolmogorov41a}
---. 1941{\natexlab{b}}, Proceedings: Mathematical and Physical Sciences, 434,
  9

\bibitem[{Kraichnan(1965)}]{kraichnan65}
Kraichnan, R.~H. 1965, Physics of Fluids, 8, 1385, \dodoi{10.1063/1.1761412}

\bibitem[{Lithwick {et~al.}(2007)Lithwick, Goldreich, \& Sridhar}]{lithwick07}
Lithwick, Y., Goldreich, P., \& Sridhar, S. 2007, The Astrophysical Journal,
  655, 269, \dodoi{10.1086/509884}

\bibitem[{MacBride {et~al.}(2005)MacBride, Forman, \& Smith}]{macbride05}
MacBride, B.~T., Forman, M.~A., \& Smith, C.~W. 2005, Solar Wind 11/SOHO 16,
  Connecting Sun and Heliosphere, 592, 613.
\newblock \url{https://ui.adsabs.harvard.edu/2005ESASP.592..613M/abstract}

\bibitem[{Mallet \& Schekochihin(2017)}]{Mallet17a}
Mallet, A., \& Schekochihin, A.~A. 2017, Monthly Notices of the Royal
  Astronomical Society, 466, 3918, \dodoi{10.1093/mnras/stw3251}

\bibitem[{Mallet {et~al.}(2016)Mallet, Schekochihin, Chandran, Chen, Horbury,
  Wicks, \& Greenan}]{mallet16}
Mallet, A., Schekochihin, A.~A., Chandran, B. D.~G., {et~al.} 2016, Monthly
  Notices of the Royal Astronomical Society, 459, 2130,
  \dodoi{10.1093/mnras/stw802}

\bibitem[{Osman {et~al.}(2014)Osman, Kiyani, Chapman, \& Hnat}]{osman14}
Osman, K.~T., Kiyani, K.~H., Chapman, S.~C., \& Hnat, B. 2014, The
  Astrophysical Journal, 783, L27, \dodoi{10.1088/2041-8205/783/2/L27}

\bibitem[{Perez \& Boldyrev(2009)}]{perez09}
Perez, J.~C., \& Boldyrev, S. 2009, Physical Review Letters, 102,
  \dodoi{\{10.1103/PhysRevLett.102.025003\}}

\bibitem[{Perez {et~al.}(2012)Perez, Mason, Boldyrev, \& Cattaneo}]{perez12}
Perez, J.~C., Mason, J., Boldyrev, S., \& Cattaneo, F. 2012, Physical Review X,
  2, 041005, \dodoi{10.1103/PhysRevX.2.041005}

\bibitem[{Politano \& Pouquet(1995)}]{politano95}
Politano, H., \& Pouquet, A. 1995, Physical Review E, 52, 636,
  \dodoi{10.1103/PhysRevE.52.636}

\bibitem[{Praskovsky \& Oncley(1994)}]{praskovsky94}
Praskovsky, A., \& Oncley, S. 1994, Phys. Rev. Lett., 73, 3399,
  \dodoi{10.1103/PhysRevLett.73.3399}

\bibitem[{Salem {et~al.}(2009)Salem, Mangeney, Bale, \& Veltri}]{salem09}
Salem, C., Mangeney, A., Bale, S.~D., \& Veltri, P. 2009, The Astrophysical
  Journal, 702, 537, \dodoi{10.1088/0004-637X/702/1/537}

\bibitem[{Savitzky \& Golay(1964)}]{savitzky64}
Savitzky, A., \& Golay, M. J.~E. 1964, Analytical Chemistry, 36, 1627,
  \dodoi{10.1021/ac60214a047}

\bibitem[{She \& Leveque(1994)}]{she94}
She, Z.-S., \& Leveque, E. 1994, Physical Review Letters, 72, 336,
  \dodoi{10.1103/PhysRevLett.72.336}

\bibitem[{Sorriso-Valvo {et~al.}(1999)Sorriso-Valvo, Carbone, Veltri,
  Consolini, \& Bruno}]{sorriso-valvo99}
Sorriso-Valvo, L., Carbone, V., Veltri, P., Consolini, G., \& Bruno, R. 1999,
  Geophysical Research Letters, 26, 1801, \dodoi{10.1029/1999GL900270}

\bibitem[{Sundkvist {et~al.}(2007)Sundkvist, Retinò, Vaivads, \&
  Bale}]{sundkvist07}
Sundkvist, D., Retinò, A., Vaivads, A., \& Bale, S.~D. 2007, Physical Review
  Letters, 99, 025004, \dodoi{10.1103/PhysRevLett.99.025004}

\bibitem[{Taylor(1938)}]{Taylor38}
Taylor, G.~I. 1938, Proceedings of the Royal Society of London Series A, 164,
  476, \dodoi{10.1098/rspa.1938.0032}

\bibitem[{Zhdankin {et~al.}(2016)Zhdankin, Boldyrev, \& Uzdensky}]{zhdankin16}
Zhdankin, V., Boldyrev, S., \& Uzdensky, D.~A. 2016, Physics of Plasmas, 23,
  055705, \dodoi{10.1063/1.4944820}

\end{thebibliography}
%\bibliographystyle{aasjournal}
%\bibliographystyle{unsrt}

\end{document}